\documentclass[%
superscriptaddress,
preprintnumbers,
 amsmath,amssymb,
 aps,
pra,
twocolumn
]{revtex4-1}

\usepackage{graphicx}
\usepackage{dcolumn}
\usepackage{bm}
\usepackage{lineno}
\usepackage{units}
\usepackage{hyperref}

\begin{abstract}
We show experimentally that an effect of motion of ions, observed in a plasma-based accelerator, depends inversely on the plasma ion mass. %
The effect appears within a single wakefield event and manifests itself as a bunch tail, occurring only when sufficient motion of ions suppresses wakefields. %
Wakefields are driven resonantly by multiple bunches, and simulation results indicate that the ponderomotive force causes the motion of ions. %
In this case, the effect is also expected to depend on the amplitude of the wakefields, experimentally confirmed through variations in the drive bunch charge. %

\end{abstract}
\begin{document}

\title{Experimental Observation of Motion of Ions in a Resonantly Driven Plasma Wakefield Accelerator}

\author{M.~Turner}
\affiliation{CERN, 1211 Geneva 23, Switzerland}
\author{E.~Walter}
\affiliation{Max Planck Institute for Plasma Physics, 85748 Garching, Germany}
\affiliation{Exzellenzcluster ORIGINS, 85748 Garching, Germany}
\author{C.~Amoedo}
\affiliation{CERN, 1211 Geneva 23, Switzerland}
\author{N.~Torrado}
\affiliation{CERN, 1211 Geneva 23, Switzerland}
\affiliation{GoLP/Instituto de Plasmas e Fus\~{a}o Nuclear, Instituto Superior T\'{e}cnico, Universidade de Lisboa, 1049-001 Lisbon, Portugal}
\author{N.~Lopes}
\affiliation{GoLP/Instituto de Plasmas e Fus\~{a}o Nuclear, Instituto Superior T\'{e}cnico, Universidade de Lisboa, 1049-001 Lisbon, Portugal}
\author{A.~Sublet}
\affiliation{CERN, 1211 Geneva 23, Switzerland}
\author{M.~Bergamaschi}
\affiliation{Max Planck Institute for Physics, 80805 Munich, Germany}
\author{J.~Pucek}
\affiliation{Max Planck Institute for Physics, 80805 Munich, Germany}
\author{J.~Mezger}
\affiliation{Max Planck Institute for Physics, 80805 Munich, Germany}
\author{N.~van Gils}
\affiliation{CERN, 1211 Geneva 23, Switzerland}
\author{L.~Verra}
\altaffiliation{Present Address: INFN Laboratori Nazionali di Frascati, 00044, Frascati, Italy}
\affiliation{CERN, 1211 Geneva 23, Switzerland}
\author{G.~Zevi Della Porta}
\affiliation{CERN, 1211 Geneva 23, Switzerland}
\affiliation{Max Planck Institute for Physics, 80805 Munich, Germany}
\author{J.~Farmer}
\affiliation{Max Planck Institute for Physics, 80805 Munich, Germany}
\author{A.~Clairembaud}
\affiliation{CERN, 1211 Geneva 23, Switzerland}
\affiliation{Max Planck Institute for Physics, 80805 Munich, Germany}
\author{F.~Pannell}
\affiliation{UCL, London WC1 6BT, United Kingdom}
\author{E.~Gschwendtner}
\affiliation{CERN, 1211 Geneva 23, Switzerland}
\author{P.~Muggli}
\affiliation{Max Planck Institute for Physics, 80805 Munich, Germany}

\collaboration{AWAKE Collaboration}

\author{R.~Agnello}
\affiliation{Ecole Polytechnique Federale de Lausanne (EPFL), Swiss Plasma Center (SPC), 1015 Lausanne, Switzerland}
\author{C.C.~Ahdida}
\affiliation{CERN, 1211 Geneva 23, Switzerland}
\author{Y.~Andrebe}
\affiliation{Ecole Polytechnique Federale de Lausanne (EPFL), Swiss Plasma Center (SPC), 1015 Lausanne, Switzerland}
\author{O.~Apsimon}
\affiliation{University of Manchester M13 9PL, Manchester M13 9PL, United Kingdom}
\affiliation{Cockcroft Institute, Warrington WA4 4AD, United Kingdom}
\author{R.~Apsimon}
\affiliation{Cockcroft Institute, Warrington WA4 4AD, United Kingdom} 
\affiliation{Lancaster University, Lancaster LA1 4YB, United Kingdom}
\author{J.M.~Arnesano}
\affiliation{CERN, 1211 Geneva 23, Switzerland}
\author{V.~Bencini}
\affiliation{CERN, 1211 Geneva 23, Switzerland}
\affiliation{John Adams Institute, Oxford University, Oxford OX1 3RH, United Kingdom}
\author{P.~Blanchard}
\affiliation{Ecole Polytechnique Federale de Lausanne (EPFL), Swiss Plasma Center (SPC), 1015 Lausanne, Switzerland}
\author{K.P.~Blum}
\affiliation{CERN, 1211 Geneva 23, Switzerland}
\author{P.N.~Burrows}
\affiliation{John Adams Institute, Oxford University, Oxford OX1 3RH, United Kingdom}
\author{B.~Buttensch{\"o}n}
\affiliation{Max Planck Institute for Plasma Physics, 17491 Greifswald, Germany}
\author{A.~Caldwell}
\affiliation{Max Planck Institute for Physics, 80805 Munich, Germany}
\author{M.~Chung}
\affiliation{POSTECH, Pohang 37673, Republic of Korea}
\author{D.A.~Cooke}
\affiliation{UCL, London WC1 6BT, United Kingdom}
\author{C.~Davut}
\affiliation{University of Manchester M13 9PL, Manchester M13 9PL, United Kingdom}
\affiliation{Cockcroft Institute, Warrington WA4 4AD, United Kingdom} 
\author{G.~Demeter}
\affiliation{HUN-REN Wigner Research Centre for Physics, Budapest, Hungary}
\author{A.C.~Dexter}
\affiliation{Cockcroft Institute, Warrington WA4 4AD, United Kingdom} 
\affiliation{Lancaster University, Lancaster LA1 4YB, United Kingdom}
\author{S.~Doebert}
\affiliation{CERN, 1211 Geneva 23, Switzerland}

\author{A.~Fasoli}
\affiliation{Ecole Polytechnique Federale de Lausanne (EPFL), Swiss Plasma Center (SPC), 1015 Lausanne, Switzerland}
\author{R.~Fonseca}
\affiliation{ISCTE - Instituto Universit\'{e}ario de Lisboa, 1049-001 Lisbon, Portugal}  
\affiliation{GoLP/Instituto de Plasmas e Fus\~{a}o Nuclear, Instituto Superior T\'{e}cnico, Universidade de Lisboa, 1049-001 Lisbon, Portugal}
\author{I.~Furno}
\affiliation{Ecole Polytechnique Federale de Lausanne (EPFL), Swiss Plasma Center (SPC), 1015 Lausanne, Switzerland}
\author{E.~Granados}
\affiliation{CERN, 1211 Geneva 23, Switzerland}
\author{M.~Granetzny}
\affiliation{University of Wisconsin, Madison, WI 53706, USA}
\author{T.~Graubner}
\affiliation{Philipps-Universit{\"a}t Marburg, 35032 Marburg, Germany}
\author{O.~Grulke}
\affiliation{Max Planck Institute for Plasma Physics, 17491 Greifswald, Germany}
\affiliation{Technical University of Denmark, 2800 Kgs. Lyngby, Denmark}
\author{E.~Guran}
\affiliation{CERN, 1211 Geneva 23, Switzerland}
\author{J.~Henderson}
\affiliation{Cockcroft Institute, Warrington WA4 4AD, United Kingdom}
\affiliation{STFC/ASTeC, Daresbury Laboratory, Warrington WA4 4AD, United Kingdom}
\author{F.~Jenko}
\affiliation{Max Planck Institute for Plasma Physics, 85748 Garching, Germany}
\author{M.A.~Kedves}
\affiliation{HUN-REN Wigner Research Centre for Physics, Budapest, Hungary}
\author{F.~Kraus}
\affiliation{Philipps-Universit{\"a}t Marburg, 35032 Marburg, Germany}
\author{M.~Krupa}
\affiliation{CERN, 1211 Geneva 23, Switzerland}
\author{T.~Lefevre}
\affiliation{CERN, 1211 Geneva 23, Switzerland}
\author{L.~Liang}
\affiliation{University of Manchester M13 9PL, Manchester M13 9PL, United Kingdom}
\affiliation{Cockcroft Institute, Warrington WA4 4AD, United Kingdom}
\author{S.~Liu}
\affiliation{TRIUMF, Vancouver, BC V6T 2A3, Canada}
\author{K.~Lotov}
\affiliation{Budker Institute of Nuclear Physics SB RAS, 630090 Novosibirsk, Russia}
\affiliation{Novosibirsk State University, 630090 Novosibirsk , Russia}
\author{M.~Martinez~Calderon}
\affiliation{CERN, 1211 Geneva 23, Switzerland}
\author{S.~Mazzoni}
\affiliation{CERN, 1211 Geneva 23, Switzerland}
\author{P.I.~Morales~Guzm\'{a}n}
\affiliation{Max Planck Institute for Physics, 80805 Munich, Germany}
\author{M.~Moreira}
\affiliation{GoLP/Instituto de Plasmas e Fus\~{a}o Nuclear, Instituto Superior T\'{e}cnico, Universidade de Lisboa, 1049-001 Lisbon, Portugal}
\author{T.~Nechaeva}
\affiliation{Max Planck Institute for Physics, 80805 Munich, Germany}
\author{N.~Okhotnikov}
\affiliation{Budker Institute of Nuclear Physics SB RAS, 630090 Novosibirsk, Russia}
\affiliation{Novosibirsk State University, 630090 Novosibirsk , Russia}
\author{C.~Pakuza}
\affiliation{John Adams Institute, Oxford University, Oxford OX1 3RH, United Kingdom}
\author{A.~Pardons}
\affiliation{CERN, 1211 Geneva 23, Switzerland}
\author{K.~Pepitone}
\affiliation{Angstrom Laboratory, Department of Physics and Astronomy, 752 37 Uppsala, Sweden}
\author{E.~Poimendidou}
\affiliation{CERN, 1211 Geneva 23, Switzerland}
\author{A.~Pukhov}
\affiliation{Heinrich-Heine-Universit{\"a}t D{\"u}sseldorf, 40225 D{\"u}sseldorf, Germany}
\author{R.L.~Ramjiawan}
\affiliation{CERN, 1211 Geneva 23, Switzerland}
\affiliation{John Adams Institute, Oxford University, Oxford OX1 3RH, United Kingdom}
\author{L.~Ranc}
\affiliation{Max Planck Institute for Physics, 80805 Munich, Germany}
\author{S.~Rey}
\affiliation{CERN, 1211 Geneva 23, Switzerland}
\author{R.~Rossel}
\affiliation{CERN, 1211 Geneva 23, Switzerland}
\author{H.~Saberi}
\affiliation{University of Manchester M13 9PL, Manchester M13 9PL, United Kingdom}
\affiliation{Cockcroft Institute, Warrington WA4 4AD, United Kingdom}
\author{O.~Schmitz}
\affiliation{University of Wisconsin, Madison, WI 53706, USA}
\author{E.~Senes}
\affiliation{CERN, 1211 Geneva 23, Switzerland}
\author{F.~Silva}
\affiliation{INESC-ID, Instituto Superior Técnico, Universidade de Lisboa, 1049-001 Lisbon, Portugal}
\author{L.~Silva}
\affiliation{GoLP/Instituto de Plasmas e Fus\~{a}o Nuclear, Instituto Superior T\'{e}cnico, Universidade de Lisboa, 1049-001 Lisbon, Portugal}
\author{B.~Spear}
\affiliation{John Adams Institute, Oxford University, Oxford OX1 3RH, United Kingdom}
\author{C.~Stollberg}
\affiliation{Ecole Polytechnique Federale de Lausanne (EPFL), Swiss Plasma Center (SPC), 1015 Lausanne, Switzerland}
\author{C.~Swain}
\affiliation{Cockcroft Institute, Warrington WA4 4AD, United Kingdom}
\affiliation{University of Liverpool, Liverpool L69 7ZE, United Kingdom}
\author{A.~Topaloudis}
\affiliation{CERN, 1211 Geneva 23, Switzerland}
\author{P.~Tuev}
\affiliation{Budker Institute of Nuclear Physics SB RAS, 630090 Novosibirsk, Russia}
\affiliation{Novosibirsk State University, 630090 Novosibirsk , Russia}
\author{F.~Velotti}
\affiliation{CERN, 1211 Geneva 23, Switzerland}
\author{V.~Verzilov}
\affiliation{TRIUMF, Vancouver, BC V6T 2A3, Canada}
\author{J.~Vieira}
\affiliation{GoLP/Instituto de Plasmas e Fus\~{a}o Nuclear, Instituto Superior T\'{e}cnico, Universidade de Lisboa, 1049-001 Lisbon, Portugal}
\author{C.~Welsch}
\affiliation{Cockcroft Institute, Warrington WA4 4AD, United Kingdom}
\affiliation{University of Liverpool, Liverpool L69 7ZE, United Kingdom}
\author{M.~Wendt}
\affiliation{CERN, 1211 Geneva 23, Switzerland}
\author{M.~Wing}
\affiliation{UCL, London WC1 6BT, United Kingdom}
\author{J.~Wolfenden}
\affiliation{Cockcroft Institute, Warrington WA4 4AD, United Kingdom}
\affiliation{University of Liverpool, Liverpool L69 7ZE, United Kingdom}
\author{B.~Woolley}
\affiliation{CERN, 1211 Geneva 23, Switzerland}
\author{G.~Xia}
\affiliation{Cockcroft Institute, Warrington WA4 4AD, United Kingdom}
\affiliation{University of Manchester M13 9PL, Manchester M13 9PL, United Kingdom}
\author{V.~Yarygova}
\affiliation{Budker Institute of Nuclear Physics SB RAS, 630090 Novosibirsk, Russia}
\affiliation{Novosibirsk State University, 630090 Novosibirsk , Russia}
\author{M.~Zepp}
\affiliation{University of Wisconsin, Madison, WI 53706, USA}
\noaffiliation

\date{\today}
\maketitle

Plasma wakefield acceleration is a novel and innovative concept for accelerating charged particles~\cite{PhysRevLett.43.267,PhysRevLett.54.693}. %
Acceleration with gradients of tens of GeV/m~\cite{PhysRevLett.122.084801,Blumenfeld2007} has been experimentally demonstrated. %
As these gradients are significantly larger than those sustained by radio-frequency cavities (\unit[$\sim$100]{MV/m}), the concept has the potential to reduce the footprint of future high-energy linear accelerators. %

In plasma wakefield accelerators, the accelerating structure is formed and sustained by plasma electrons, which oscillate collectively (with a plasma period $\tau_{pe}$~\cite{taupe}) in a background of positively-charged plasma ions (often assumed to be immobile and uniformly distributed). %
Most commonly, wakefields are excited by a single, short and dense or intense driver (relativistic charged particle bunch or laser pulse) fitting within $\tau_{pe}$. %
However, wakefields can also be excited resonantly by a train of less dense or intense drivers spaced at $\tau_{pe}$. %
Such a driver train can be preformed~\cite{muggli,hooker,doi:10.1142/S0217979207042197} or be the result of a self-modulation process~\cite{lasersm1,lasersm2,PhysRevLett.104.255003}. %
In both schemes, a properly placed charged particle bunch (witness) is accelerated and transversely focused by the wakefields. %

With a single driver and the accelerated bunch in the same period of the wakefields, the accelerator usually operates in the blow-out regime~\cite{Pukhov2002}. %
In this case, a negatively-charged witness bunch travels in the uniform ion column left behind the driver, which provides an ideal restoring force profile, i.e., increasing linearly with distance from the axis. %
However, the force profile is modified when motion of ions, e.g. caused by the response to the impulse force from the fields of the driver or of the accelerated bunch, leads to a non-uniform ion density distribution within one period of the wakefields. %
This leads to witness bunch emittance growth, which may be unacceptable, particularly in the context of a collider~\cite{rosenzweig}. However, motion of ions has also been proposed as a beneficial mechanism, specifically to suppress beam-hose instability~\cite{PhysRevLett.67.991,PhysRevLett.132.075001}, which compromises the acceleration process~\cite{Balakin:1983sc,mehrling} and may impose a fundamental limitation on acceleration efficiency~\cite{PhysRevAccelBeams.20.121301}. %
In this scenario, motion of ions serves as a suppression mechanism, similar to the Balakin-Novokhatski-Smirnov (BNS) damping~\cite{Novokhatski:2020kyl,Balakin:1983sc}. %
BNS damping is an established technique for enhancing the performance of conventional linear accelerators, e.g. for increasing their luminosity~\cite{Assmann:1997se}. %

When using a train of drivers to excite wakefields resonantly, the witness bunch is placed not in the first, but in the $n^{th}$ ($n>1$) period of the wakefields, since the amplitude of the wakefields grows along the train. %
Therefore, the effect of the motion of ions over $n$ periods must be considered. %
A new cause for the motion of ions may become dominant, i.e., the cumulative effect of the ponderomotive force of the wakefields themselves acting on the ions~\cite{10.1063/1.1559011,PhysRevLett.109.145005,10.1063/1.4876620}. %
In this case, the motion of ions indirectly perturbs the acceleration process by perturbing the driving of the wakefields. %

The expression for the ponderomotive force of a plasma wave of angular frequency $\omega=\omega_{pe}$ is~\cite{PhysRevLett.109.145005}:
\begin{equation}
    \label{eq:1}
    \boldsymbol{F}_p\cong-\frac{e^2}{4m_e\omega_{pe}^2}\boldsymbol{\nabla} \tilde{W_r}^2,
    \end{equation}
where $e$ is the electron charge, $m_e$ is the electron mass, $\omega_{pe}=\sqrt{\frac{n_{pe} e^2}{\epsilon_0 m_e}}$ is the plasma electron angular frequency, $n_{pe}$ is the plasma electron density, $\epsilon_0$ is the vacuum dielectric constant and $\tilde{W_r}$ is the envelope of the transverse wakefield. %
Eq.~\ref{eq:1} shows that $|\boldsymbol{F}_p|$ scales with the square of the envelope of the transverse wakefields $\tilde{W_r}$ (for a fixed transverse bunch size $\sigma_r$, as $W_r$ is zero on-axis and increases radially to a maximum value over a distance of approximately $\sigma_r$). %
The effect of the ponderomotive force on the ion density was evidenced with a shadowgraphy diagnostic~\cite{PhysRevX.9.011046} following a single electron bunch driving wakefields. %

Using theory and simulations, the effects of $\boldsymbol{F}_p$ and its dependencies can be quantified. 
In Ref.~\cite{lotovion}, it was found that the time along the bunch for wavebreaking to occur as a result of motion of ions scales as $m_i^{-1/3}$, where $m_i$ is the ion mass. 
We note here that the expected inverse relationship with plasma ion mass is common to all effects caused by motion of ions (e.g. resonance detuning, emittance growth...) and regardless of the force moving the ions, through Newton's equation. %

When a microbunch train is formed by self-modulation, it was shown in theory and simulations~\cite{PhysRevLett.109.145005,10.1063/1.4876620} that the motion of ions leads to the crossing of plasma electron trajectories~\cite{PhysRev.113.383} late along the bunch train and plasma. %
This results in a loss of coherence in the collective motion of plasma electrons and therefore to a decrease in wakefield amplitude that imprints itself on the bunch during the self-modulation process. %

In this \emph{Letter}, we demonstrate experimentally for the first time that an effect of motion of ions observed in a plasma-based accelerator depends on the mass of the plasma ions. %
The effect appears within a single wakefield event and, as expected, first with lighter ions, all other parameters kept equal. %
Since wakefields are driven by multiple bunches, simulation results indicate that the ponderomotive force causes the motion of ions. %
In this case, the effect is also expected to scale with the amplitude of the wakefields, also confirmed by our observations. %
When the motion of ions becomes significant, wakefields and the formation of the microbunch train are suppressed. %
This suppression manifests itself as a 'tail'—a late increase in density—in time-resolved images of the bunch, as previously predicted by simulation studies~\cite{PhysRevLett.109.145005,10.1063/1.4876620}. %
Since the tail is caused by decoherence of electron motion, or wavebreaking, the effect scales with $m_i^{-1/3}$, as in Ref.~\cite{lotovion}.

\begin{figure}[htb!]
    \centering
    \includegraphics[width=0.5\textwidth]{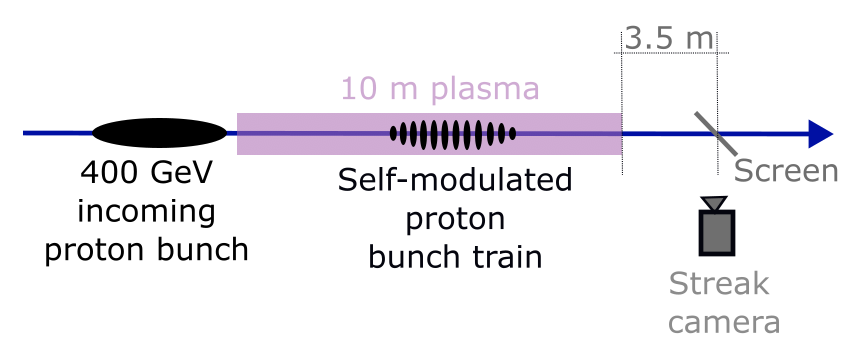}
    \caption{Schematic of the experimental setup. %
    }
    \label{fig:schematic}
\end{figure}

Measurements are performed at AWAKE (Advanced WAKefield Experiment) at CERN. %
A schematic layout of the experiment is shown in Fig.~\ref{fig:schematic}. %
AWAKE uses a bunch of \unit[400]{GeV} protons from the Super Proton Synchrotron (SPS) to drive wakefields over \unit[10]{m} of plasma. %
The bunch contains $N_{p^+}=$(0.7-3)$\times$10$^{11}$ protons, is transversely focused to an rms size of \unit[$\sigma_{\mathrm{x_0,y_0}}\approx(160\pm4)$]{$\mu$m} near the plasma entrance, and has an rms duration of \unit[$\sigma_{\xi}\approx(170\pm2)$]{ps}. %
The bunch is much longer than $\tau_\mathrm{pe}=2\pi/\omega_{pe}\cong $\unit[(10-3)]{ps} (\unit[$n_{pe}\cong(1-10) \times 10^{14}$]{cm$^{-3}$}, typical of these experiments) and undergoes the self-modulation instability (SMI) over the first few meters of plasma~\cite{PhysRevLett.122.054802,PhysRevLett.122.054801,PhysRevAccelBeams.23.081302}. %
The SMI results in the formation of a periodic microbunch train with a spacing of $\approx\tau_\mathrm{pe}$. %
This train drives wakefields resonantly along the bunch and plasma, producing large amplitude wakefields. %

The plasma is provided by a pulsed DC discharge source~\cite{10355763,carolina} and is either made of helium ($^4$He), argon ($^{40}$Ar) or xenon ($^{131}$Xe~\cite{Xecomposition}) ($m_{\mathrm{Ar}} \cong 10\times m_{\mathrm{He}}$ and $m_{\mathrm{Xe}} \cong 3\times m_{\mathrm{Ar}}$). %
The discharge source has \unit[0.2]{mm} aluminum windows at its entrance (and exit) that exclude the option of seeding with a relativistic ionization front~\cite{PhysRevLett.122.054802,PhysRevLett.126.164802} or with a low-energy electron bunch~\cite{PhysRevLett.129.024802}. %
Therefore, in these experiments, self-modulation grows from noise as an instability. %

Gases are only partially and at most singly ionized~\cite{carolina}. %
The plasma density is adjusted by changing the gas pressure (\unit[8 to 45]{Pa}), the peak discharge current (\unit[300 to 600]{A}, pulse duration \unit[$\approx$ 25]{$\mu$s}), and the timing between the discharge and the arrival time of the proton bunch, so that similar plasma densities can be reached with different gases. %
Reachable plasma density ranges are \unit[$n_\mathrm{pe}$=(0.1-4.8)$\times10^{14}$]{cm$^{-3}$} with helium, \unit[$n_\mathrm{pe}$=(0.1-10)$\times10^{14}$]{cm$^{-3}$} with argon, and \unit[$n_\mathrm{pe}$=(1-17)$\times10^{14}$]{cm$^{-3}$} with xenon. %
Plasma densities are measured either by longitudinal, double-pass interferometry (prior to the experiment) or by measuring the modulation frequency of the microbunch trains resulting from SMI~\cite{PhysRevLett.122.054802,PhysRevLett.120.144802,carolina}, and are averages over the plasma length~\footnote{The density values obtained from the two methods show a consistent difference of approximately \unit[10]{\%}.}. %

At a distance of \unit[3.5]{m} downstream of the plasma exit, protons traverse a screen (\unit[150]{$\mu$m} thick SiO\textsubscript{2}, aluminum coated) and emit transition radiation (Fig.~\ref{fig:schematic}). %
Wavelengths in the \unit[(450$\pm$25)]{nm} range are imaged onto the entrance slit of a streak camera, that provides time-resolved images of the bunch density distribution in a \unit[$\Delta y=$80]{$\mu$m}-wide slice around its axis~\footnote{For the streak camera settings used in this manuscript, the minimum detectable signal corresponds to approximately \unit[0.25]{pC/(ps mm)}.}. %

\begin{figure}[htb!]
    \centering
    \includegraphics[width=0.5\textwidth]{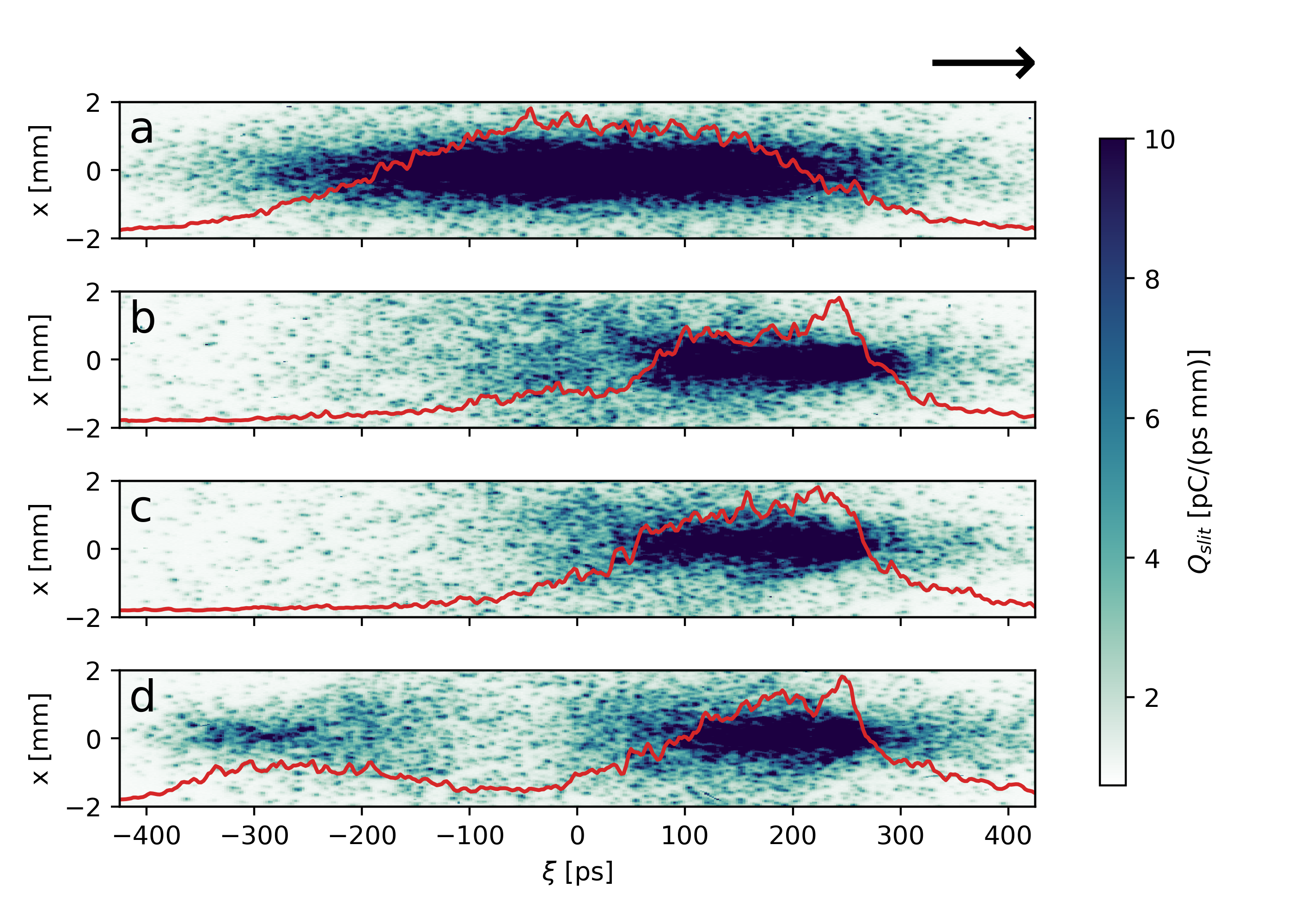}
    \caption{Single bunch measurements of the time-resolved proton bunch density $n_{p^+}(x,\xi)$ with $N_{p^+}$=(2.8$\pm$0.1)$\times$10$^{11}$ measured \unit[3.5]{m} downstream of the plasma exit without plasma (a) and with \unit[10]{m} of \unit[$n_{pe}$=(4.8$\pm$0.2)$\times$10$^{14}$]{cm$^{-3}$} xenon (b), argon (c) and helium (d) plasmas. %
   Red lines show the vertical sum. %
    The longitudinal bunch center is at $\xi=0$. %
    Bunches propagate to the right as indicated by the arrow on the top right. %
    Color-scale saturated to highlight the bunch tail. %
    Identical streak camera settings used for all measurements. %
    }
    \label{fig:nsimgs}
\end{figure}

Figure~\ref{fig:nsimgs}a shows the %
time-resolved measured proton bunch density $n_{p^+}(x,\xi)$ with $N_{p^+}$=(2.8$\pm$0.1)$\times$10$^{11}$ after propagation in vacuum (no plasma). %
The distribution is approximately bi-Gaussian with a transverse rms size of \unit[$\sigma_{\mathrm{x,y,SC}}=625$]{$\mu$m} ($\gg \Delta y$). %
Figure~\ref{fig:nsimgs} shows the density after propagating in xenon (b) and argon (c) plasmas with \unit[$n_\mathrm{pe}$=(4.8$\pm$0.2)$\times$10$^{14}$]{cm$^{-3}$}, the highest density reachable with helium. %
These show typical features of successful SMI~\cite{PhysRevLett.122.054802,PhysRevLett.122.054801}:
observable microbunch structure when using shorter time-windows (\unit[73]{ps}, Supplemental Material~\cite{SupplementalMaterial})~\cite{PhysRevLett.122.054802}; decrease of the transverse bunch size (visible from the front of the bunch to \unit[$\xi\simeq$ 250]{ps}), caused by adiabatic focusing of the bunch in plasma~\cite{verra2022adiabatic}; signal decrease (visible for \unit[$\xi\lesssim 200$]{ps}, when compared to the no-plasma case on Fig.~\ref{fig:nsimgs}a), caused by the increase of proton divergence along the bunch~\cite{EAAC2023SC}. %
The divergence increases because the transverse wakefield amplitude increases due to resonant wakefield excitation. %
Protons with larger transverse momentum diverge more during vacuum propagation downstream of the plasma exit, leading to a lower bunch density measured with the streak camera because of the effect of the slit. %
Very little to no signal is observed for \unit[$\xi\lesssim$ -100]{ps} on Figs.~\ref{fig:nsimgs}b,c.

\begin{figure*}[htb!]
    \centering
    \includegraphics[width=\textwidth]{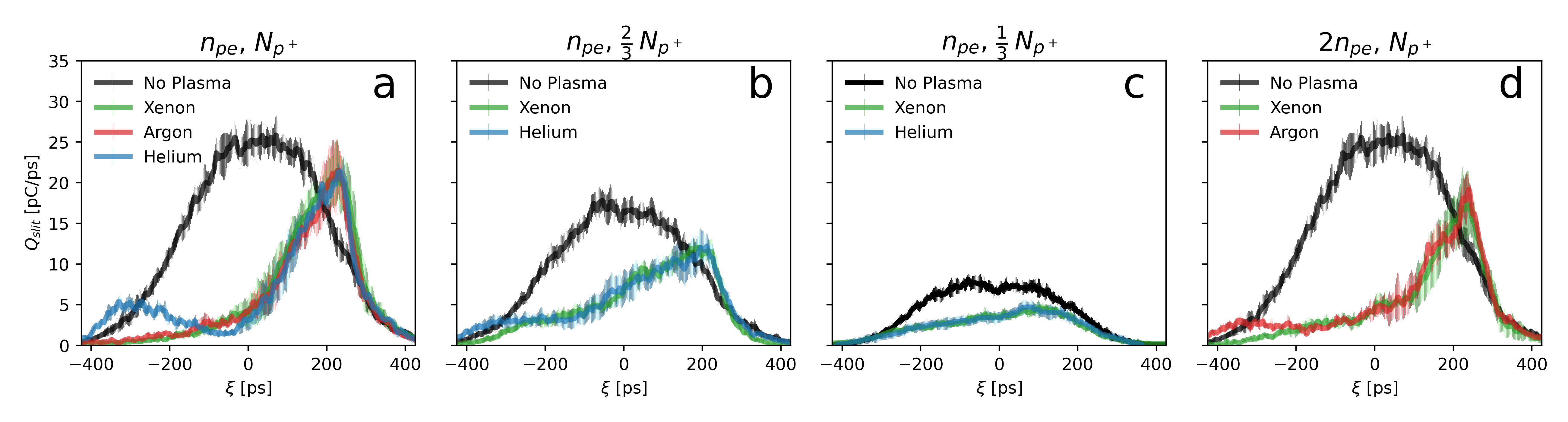}
    \caption{Bunch density profiles $n_{p^+}(\xi)$ of time-resolved images using the range \unit[$|x|<$0.75]{mm} of the $n_{p^+}(x,\xi)$ measurements. %
    Each line represents the average of typically ten measurements. %
    The standard deviation is shown by vertical error-bars. %
    Bunches propagate to the right. %
    Bunch and plasma parameters given in the titles and labels, with \unit[$n_{pe}=(4.8\pm0.2)\times10^{14}$]{cm$^{-3}$} and $N_{p^+}=(2.8\pm0.1)\times10^{11}$. %
    }
    \label{fig:lineouts}
\end{figure*}

The bunch density measured with helium (Fig.~\ref{fig:nsimgs}d) closely resembles those with argon (Fig.~\ref{fig:nsimgs}c) or xenon (Fig.~\ref{fig:nsimgs}b) from the front of the bunch to \unit[$\xi\sim -100$]{ps}, showing essentially the same SMI development and wakefield growth along the bunch and plasma. %
Microbunches are visible with all three plasmas on shorter time-windows also in this case (Supplemental Material~\cite{SupplementalMaterial}). %
However with helium (Fig.~\ref{fig:nsimgs}d) and for \unit[$\xi\lesssim-100$]{ps}, the bunch density increases again, leading to the appearance of a bunch tail, not present on Figs.~\ref{fig:nsimgs}b and c~\footnote{The intensity of the bunch tail around \unit[$\xi \sim -300$]{ps} is higher than when no plasma is present, due to adiabatic focusing~\cite{PhysRevLett.129.024802, verra2022adiabatic}.}. %
This is the signature expected from the effect of the motion of ions on self-modulation~\cite{PhysRevLett.109.145005,10.1063/1.4876620}. %
This occurs when the mass of the ions is sufficiently low and the amplitude of the wakefields is sufficiently high for motion of ions to suppress the wakefields in the back of the bunch. %
Where wakefields are suppressed, the SMI development is suppressed and protons acquire much less transverse momentum, leading to much smaller divergence and increased proton bunch density, visible as a tail on the time-resolved image (Fig.~\ref{fig:nsimgs}d). %
The expected inverse dependence with $m_i$ is confirmed, as, with this amplitude of wakefields, only the bunch density measured with the lightest ions (helium) is disturbed. %
Simultaneous measurements with two streak cameras with orthogonal slits recording $n_{p^+}(x,\xi)$ and $n_{p^+}(y,\xi)$ (Supplemental Material~\cite{SupplementalMaterial}) confirm that the core of the bunch and its tail are radially symmetric, as expected. %

The influence of motion of ions on self-modulation is also evident from bunch density profiles $n_{p^+}(\xi)$, presented in Fig.~\ref{fig:lineouts} as averages of typically ten measurements with their standard deviation. %
Figure~\ref{fig:lineouts}a displays profiles corresponding to the four images on Fig.~\ref{fig:nsimgs}. %
These profiles again show the focusing effect in the front (\unit[$\xi\gtrsim$250]{ps}), i.e., higher densities with plasma compared to without plasma (black line), as well as the presence of a clear tail in the distribution observed with helium (blue line \unit[$\xi\lesssim$-100]{ps}). %
The profiles highlight again the similarity of the bunch densities with all three plasmas between the front of the bunch and \unit[$\xi\sim-100$]{ps}. %
In that range, the development of self-modulation is primarily influenced by the response of the plasma electrons. %

Figure~\ref{fig:lineouts}b shows that with helium (blue line) and $\frac{2}{3}N_{p^+}$, i.e. a lower peak wakefield amplitude (decrease by approximately 1/3 in numerical simulation~\cite{Simsimmobile}; numerical simulations detailed later) and thus less motion of ions expected, the size of the tail is reduced when compared to that with $N_{p^+}$. %
Figure~\ref{fig:lineouts}c shows that no tail is measurable with $\frac{1}{3}N_{p^+}$, i.e., with even lower wakefield amplitude (peak field in simulations approximately half of that with $N_{p^+}$). %
For Figs.~\ref{fig:lineouts}b,c, we do not plot the lines obtained with argon, since there is no measurable difference with the ones with xenon (as is the case on Fig.~\ref{fig:lineouts}a). %
These show that, for sufficiently low wakefield amplitude, achieved by decreasing $N_{p^+}$, the effect of motion of ions disappears even with the lightest ions. %

Figure~\ref{fig:lineouts}d shows that when increasing $n_{pe}$ (by approximately a factor of two), the effect of motion of ions becomes observable with $N_{p^+}$ and argon (red line, $m_\mathrm{Ar}\cong10\times m_\mathrm{He}$). %
This is because of the higher amplitude of the wakefields (increase by \unit[$\sim60$]{\%} in simulations compared to the ones with \unit[$n_{pe}=4.8\times10^{14}$]{cm$^{-3}$}) and the shorter $\tau_{pe}$ (earlier decoherence for shorter $\tau_{pe}$). %
We note that such a high density was not reachable with helium within the safe current limit of the pulse generator ~\cite{10355763}. %
The effect observed with argon is similar to that observed at lower density using helium, showing that the same physics is at play. %
The fact that we still do not observe a tail with xenon (Fig.~\ref{fig:lineouts}d, green line) shows that the amplitude of wakefields at this plasma density is now sufficient make the bunch tail observable with argon (Fig.~\ref{fig:lineouts}d, purple line), but not for xenon plasma ($m_{\mathrm{Xe}}\cong 3\times m_{\mathrm{Ar}}$). %

\begin{figure}[htb!]
    \centering
    \includegraphics[width=0.5\textwidth]{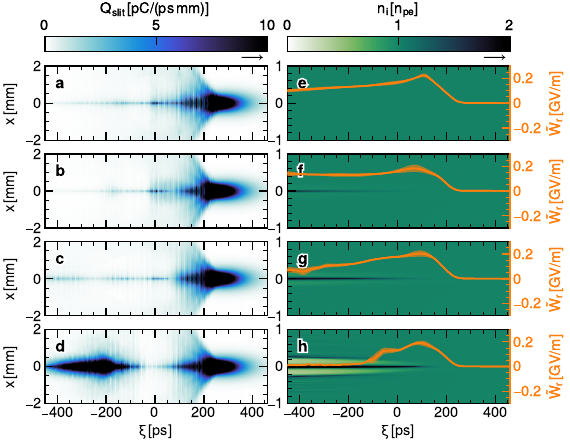}
    \caption{Numerical simulation results (average of five simulations with different noise in the bunch particle distribution) with experimental parameters from Fig.~\ref{fig:nsimgs} (\unit[$n_{pe}=4.8\times10^{14}$]{cm$^{-3}$} and $N_{p^+}=2.8\times10^{11}$) . %
    Figures a-d: bunch distributions at the location of the streak camera screen and with: immobile (a), xenon (b), argon (c), and helium (d) ions. %
    Effect of the streak-camera slit included. %
    Figures e-h: corresponding densities of plasma ions at the plasma exit ($z$=\unit[10]{m}). %
    Orange lines: envelope of the average transverse wakefields ($W_\mathrm{r}$) within $c/\omega_{pe}$, also at $z$=\unit[10]{m}. %
    Error-bars: standard deviation of the simulation results. %
Numerical simulation result with argon and \unit[$n_{pe}=9.3\times10^{14}$]{cm$^{-3}$} ($2n_{pe}$) in Supplemental Material~\cite{SupplementalMaterial}. %
    }    
    \label{fig:sim}
\end{figure}

To confirm that the observed bunch tails result from the motion of ions~\cite{10.1063/1.4876620}, we conduct particle-in-cell simulations using 2D-cylindrical LCODE~\cite{LCODE,LCODE2} with input bunch (bi-Gaussian proton bunch distribution) and plasma parameters from Fig.~\ref{fig:nsimgs}. %
The simulated proton bunch distributions are propagated in vacuum from the plasma exit to the location of the screen in the experiment (see Fig.~\ref{fig:schematic}). %
On Fig.~\ref{fig:sim}, we display the density distributions of a $\Delta y$-wide slice (width of the camera slit) around the bunch axis for immobile ions (a) and mobile ions of xenon (b), argon (b) and helium (c). %

The SMI develops from noise in the randomly initiated drive bunch distribution, leading to variations in the results from simulation to simulation. %
For the density plots of Fig.~\ref{fig:sim} we show averages of five simulation results for different numerical seeds. %
For the amplitude of the transverse wakefields, we show the peak field values within one oscillation (envelope, orange lines) and standard deviations as error bars. %
As with experimental results (Fig.~\ref{fig:nsimgs}), we observe a clear bunch tail only with helium. %
Otherwise, distributions exhibit only subtle differences, which would be difficult to distinguish in experiments due to the finite temporal and spatial resolutions of the experimental images~\cite{EAAC2023SC}. %
Simulated density profiles generally show shorter bunch fronts and higher density and longer bunch tails than experimental ones (Figs.~\ref{fig:nsimgs}d). This difference reflects stronger wakefield growth in the simulations, where the initial seed wakefield amplitudes are larger than in the experiment due to the necessary use of fewer macroparticles compared to the number of protons in the experiment. %

Nevertheless, profiles retain the same features and given the striking similarity between the bunch density distributions observed in simulations and experiments, we can deduce the underlying cause for the formation of the bunch tail from simulation results. %

With mobile ions, ion density distributions (Figs.~\ref{fig:sim}f-h) exhibit typical features caused by the ponderomotive force of wakefields driven by a narrow bunch~\cite{PhysRevLett.109.145005, 10.1063/1.1559011}, rather than from the impulse response to the bunch. %
Since transverse wakefields are axially-symmetric and peak near $\sigma_{x_0,y_0}$ from the axis, their ponderomotive force creates a high ion density region near the axis (\unit[$x\sim0$]{mm}) and a low density surrounding it, especially visible with helium (Fig.~\ref{fig:sim}h). %
This ion density perturbation changes the local plasma electron oscillation period, leading to a loss of coherence in their collective motion. %
The loss of coherence leads to a decrease of the wakefield amplitude in the back of the bunch (\unit[$\xi\lesssim-50$]{ps} on Fig.~\ref{fig:sim}h), which stops the formation of the microbunch train and results in the appearance of a bunch 'tail'~\cite{PhysRevLett.109.145005, 10.1063/1.1559011}. %

The ion density perturbation is visible even with the heaviest ions (xenon, Fig.~\ref{fig:sim}f), but becomes noticeable only very late along the bunch. %
Since in this case the relative density perturbation remains small, we observe no significant effect on the outcome of the self-modulation process. %
The effect occurs earlier and is larger with argon ions, but is too small for a significant tail to form. %

Numerical simulation results also show that the time for the tail to form along the bunch scales with $m_i^{-1/3}$ (Supplemental Material~\cite{SupplementalMaterial}). %
This scaling is logical, as the tail formation is due to the loss of coherence in plasma electron motion or wavebreaking, which has been demonstrated to scale with $m_i^{-1/3}$ in Ref.~\cite{lotovion}. %
All experimental results obtained are consistent with this scaling.%

In the absence of significant motion of ions (immobile ions and xenon Fig~\ref{fig:sim}e,f) the wakefields maintain an approximately constant amplitude after saturation (\unit[$\xi\lesssim-100$]{ps}).
With argon, the motion of ions is sufficient to cause a decrease in the wakefield amplitude for \unit[$\xi\lesssim-100$]{ps}, but not sufficient for a tail to form. %
With helium the wakefield amplitude plummets around \unit[$\xi\lesssim-50$]{ps} and thus a tail forms. %

The slow decrease in wakefield amplitude observed with immobile ions later than \unit[$\xi\sim+100$]{ps} is due to the non-optimal evolution of self-modulation in a uniform plasma~\cite{PhysRevLett.107.145002,PhysRevLett.104.255003}. %
This evolution also leads to the relatively small peak wakefield amplitudes (\unit[$\sim200$]{MV/m}) at \unit[10]{m}, from \unit[$\sim600$]{MV/m} at \unit[6]{m}. %
Numerical simulation results suggest that with a plasma density step placed early along the plasma, wakefields maintain an amplitude close to their peak value~\cite{10.1063/1.4933129,sym14081680}. %
Interestingly, the small ion density perturbation observed with xenon (Fig.~\ref{fig:sim}f) has a positive effect on the amplitude of the wakefields~\cite{lotovion,Minakov_2019,10.1063/5.0197176}. In this case, changes in the plasma electron oscillation period counteract the wakefield phase velocity shifts that arise during the development of self-modulation. %

In AWAKE, the plasma typically consists of rubidium, and both simulation and experimental results indicate that the mass of rubidium ions ($m_{\mathrm{Rb}}>2\times m_{\mathrm{Ar}}$) is sufficiently large to prevent motion of ions from having negative effects under any anticipated experimental conditions. %
Additionally, no bunch tails, such as those observed in these experiments, were ever seen with rubidium in previous experiments, even at densities as high as \unit[9.9$\times 10^{14}$]{cm$^{-3}$}~\footnote{From the transverse size of the driver, the density optimum for acceleration in AWAKE is \unit[7$\times 10^{14}$]{cm$^{-3}$}.}. %
From the presented experimental study, we can also establish 'safe limits'—defined as the clear absence of bunch tails—of \unit[1$\times 10^{14}$]{cm$^{-3}$} with helium, \unit[4$\times 10^{14}$]{cm$^{-3}$} with argon, and no upper limit was observed for xenon. %
However, most importantly, the appearance of beam tails will serve as a clear diagnostic method in all future experiments. %
In acceleration experiments, the witness bunch is positioned at the point along the bunch where wakefields reach their maximum amplitude—this would be \unit[$\xi \simeq 0$]{ps} (as shown in Fig.~\ref{fig:sim}f,g), though it may be positioned further along the bunch when, for example, density steps are used~\cite{10.1063/1.4933129}. %
Any motion of ions affecting wakefields beyond the maximum amplitude point is irrelevant to the acceleration process. %

We also note that the long-timescale (\unit[$>1$]{ns}) effect of the motion of plasma ions extends beyond that on bunch emittance and energy gain. %
It potentially imposes a limitation on the repetition rate of the acceleration process due to the time it takes for the energy of wakefields to dissipate~\cite{Zgadzaj2020} and correspondingly, the time the plasma takes to recover from the excitation-acceleration process that leaves energy in the plasma. %
This recovery time includes reaching again a uniform plasma density, i.e., uniform electron, ion, and neutral densities, as measured in Ref.~\cite{D’Arcy2022}. %

The combination of experimental and simulation results presented in this \emph{Letter} clearly show an effect of the motion of ions within a single wakefield event and for the first time with different ion masses. %
This effect (formation of a bunch tail) is caused by the ponderomotive force of the wakefields. %
The experimental results confirm that the effect depends inversely on the mass of the ions (helium, argon, xenon), i.e. appears first with lighter ions, and increases with the amplitude of the wakefields. %
The dependence observed with mass, and that with amplitude, are in agreement with those of theoretical and simulation models~\cite{PhysRevLett.109.145005,  10.1063/1.4876620,lotovion}. %
The scaling with $m_i^{-1/3}$ can be used to evaluate the possible effect of the motion of ions, e.g. in single and multiple drivers wakefield accelerators that may use light ions to avoid multiple ionization levels in the strong fields of the intense driver and witness beams. %

\section{Acknowledgements}
This work was supported in parts by Fundação para a Ciência e Tecnologia - Portugal (Nos.\ CERN/FIS-TEC/0017/2019, CERN/FIS-TEC/0034/2021, UIBD/50021/2020), STFC (AWAKE-UK, Cockcroft Institute core, John Adams Institute core, and UCL consolidated grants), United Kingdom, the National Research Foundation of Korea (Nos.\ NRF-2016R1A5A1013277 and NRF-2020R1A2C1010835).
M. W. acknowledges the support of DESY, Hamburg.
Support of the Wigner Datacenter Cloud facility through the Awakelaser project is acknowledged.
TRIUMF contribution is supported by NSERC of Canada.
UW Madison acknowledges support by NSF award PHY-1903316.
The AWAKE collaboration acknowledges the SPS team for their excellent proton delivery.

\bibliographystyle{ieeetr}
\bibliography{refs} 

\begin{thebibliography}{10}
\bibitem{Simsimmobile}
Simulations use immobile ions and therefore represent peak wakefield amplitudes without motion of ions.
\end{thebibliography}
\end{document}